\documentclass[sigconf]{acmart}

\AtBeginDocument{%
  \providecommand\BibTeX{{%
    \normalfont B\kern-0.5em{\scshape i\kern-0.25em b}\kern-0.8em\TeX}}}

\copyrightyear{2022}
\acmYear{2022}
\setcopyright{acmlicensed}\acmConference[CHI '22]{CHI Conference on Human Factors in Computing Systems}{April 29-May 5, 2022}{New Orleans, LA, USA}
\acmBooktitle{CHI Conference on Human Factors in Computing Systems (CHI '22), April 29-May 5, 2022, New Orleans, LA, USA}
\acmPrice{15.00}
\acmDOI{10.1145/3491102.3517525}
\acmISBN{978-1-4503-9157-3/22/04}

\acmSubmissionID{2068}

\usepackage{stfloats}
\usepackage{microtype}
\usepackage{multirow}
\usepackage{xcolor}

\newcommand{\modified}[1]{\textcolor{black}{#1}}

\begin{document}

\title{Polite or Direct? Conversation Design of a Smart Display for Older Adults Based on Politeness Theory}


\author{Yaxin Hu, Yuxiao Qu, Adam Maus, and Bilge Mutlu}
\affiliation{%
  \institution{University of Wisconsin--Madison}
  \streetaddress{CS Building, University of Wisconsin-Madison, Madison, WI 53706}
  \city{Madison}
  \state{WI}
  \country{USA}
  }
\email{{yaxin.hu,qu45,amaus,bmutlu}@wisc.edu}

\begin{abstract}
Conversational interfaces increasingly rely on human-like dialogue to offer a natural experience. However, relying on dialogue involving multiple exchanges for even simple tasks can overburden users, particularly older adults. In this paper, we explored the use of politeness theory in conversation design to alleviate this burden and improve user experience. To achieve this goal, we categorized the voice interaction offered by a smart display application designed for older adults into seven major speech acts: request, suggest, instruct, comment, welcome, farewell, and repair. We identified face needs for each speech act,  applied politeness strategies that best address these needs, and tested the ability of these strategies to shape the perceived politeness of a voice assistant in an online study ($n=64$). Based on the findings of this study, we designed \textit{direct} and \textit{polite} versions of the system and conducted a field study ($n=15$) in which participants used each of the versions for five days at their homes. Based on five factors merged from our qualitative findings, we identified four distinctive user personas---\textit{socially oriented follower}, \textit{socially oriented leader}, \textit{utility oriented follower}, and \textit{utility oriented leader}---that can inform personalized design of smart displays.

\end{abstract}


\begin{teaserfigure}
\centering
    \vspace{6pt}
  \includegraphics[width=\textwidth]{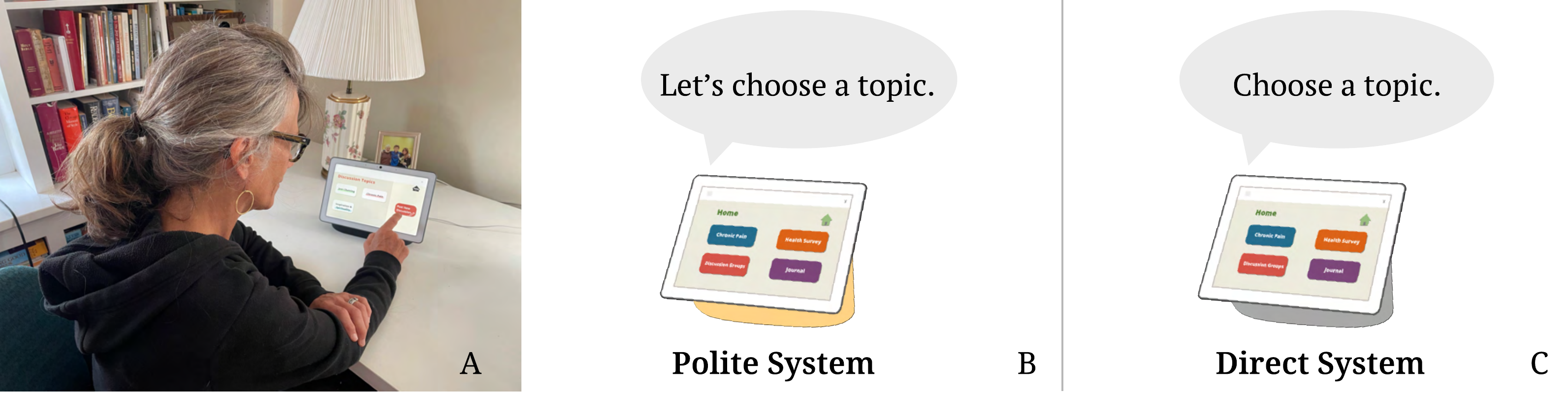}
  \caption{In this paper, we applied politeness strategies to seven speech acts of a smart display, identified four personas representing older adults' experience of system usage and preferences for politeness, and suggested personalized design strategies for each persona. (A) An older adult using the conversational interface; (B) an example of \textit{polite} speech used by our interface; and (C) an example of \textit{direct} speech used by our interface.}
  \Description{}
  \label{fig:teaser}
  \vspace{6pt}
\end{teaserfigure}

\maketitle

\section{Introduction}


\modified{
Conversational devices are increasingly being adopted in day-to-day environments, and the interaction design of the conversational behaviors of these devices is key to improving user experience with the services delivered through them. However, the design space for these behaviors involves competing objectives with significant design tradeoffs. For example, on the one hand, a major focus of conversational design guidelines proposed in prior research is the \textit{efficiency} of interaction \cite{10.1145/3411764.3445312}. Researchers have suggested that efficiency can be achieved by maintaining minimal information exchange in short utterances and making progress in the dialogue through follow-up questions \cite{moore2019conversational, fischer2019progressivity, 8383654}. On the other hand, recent work has proposed the need for proactive behavior in voice interaction design and for conversational systems to play a more active role in engaging users \cite{10.1145/3411810, 10.1145/3469595.3469618, 10.1145/3494965}. This work has demonstrated multiple rounds of conversational exchanges, directed by either the user and the device, to increase the perceived agency of the system. Designers who wish to achieve a level of conversational efficiency while following conversational norms must adopt design principles that balance both objectives of efficiency and conversational norms, although little is known about how such balance might be achieved. In this paper, we explore how \textit{linguistic politeness strategies} might inform the design of conversational interfaces toward achieving this design goal.
}

\modified{
Our work focuses on older adults as the target user group as many of them face motor and vision issues, and conversational interfaces hold great potential to address these accessibility challenges. The main premise of our work is that polite speech can meet expectations of conversational norms while maintaining a level of conversational efficiency. Therefore, we explore how a conversational interface can adopt a \textit{direct} speech style as opposed to a \textit{polite} speech style in its interactions with users.} The notions of \textit{directness} and \textit{politeness} in language have been characterized by two linguistic models: Grice's Maxims \cite{grice1975logic} and Politeness Theory \cite{brown_politeness_1987}. The \emph{manner} dimension of Grice's Maxims emphasizes clarity and brevity for effective information exchange, whereas Politeness Theory considers people's \textit{face} needs, i.e., the public self-image that people want to claim, sometimes at the cost of efficiency. Four strategies are proposed in Politeness Theory: \textit{bald-on-record}, \textit{positive politeness}, \textit{negative politeness}, and \textit{off-record}. \modified{Bald-on-record follows Grice's Maxims of manner and is characterized by directness. Politeness-based strategies, on the other hand, consider people's social needs and attend to their face needs, i.e., the desire to be appreciated and approved (positive face) and the desire to be free from imposition and distraction (negative face). Our work adopts these constructs to establish \textit{polite} and \textit{direct} speech for the conversational interface of a smart display application designed for older adults.}

\modified{
To explore the use of these linguistic politeness strategies to achieve directness and politeness in conversation design, we first categorized dialogue between the user and the conversational interface into a set of speech acts. In our smart display application, seven speech acts were identified: \textit{request}, \textit{suggest}, \textit{instruct}, \textit{comment}, \textit{welcome}, \textit{farewell}, and \textit{repair}. Next, we determined conversational goals for each speech act in terms of the user's face needs. For example, the user's face needs may be desirable when they hear the system's \emph{welcome} speech act, while the needs for \emph{request} may be to minimize imposition by the conversational interface. We selected politeness strategies that meet the face needs of users in the context of each speech act.}

We conducted an online formative study ($n=64$) to assess the ability of our strategies to shape the perceived directness and politeness of the voice assistant. Based on the findings from the formative study, we implemented two systems: a \textit{polite} version that utilized positive and negative politeness strategies, and a \textit{direct} version that utilized the bald-on-record strategy. We conducted a field study ($n=15$) to compare user experience and usage patterns across the two versions of the system. The findings revealed five factors affecting user preferences, experience, and interest in long-term adoption of the system, which led to the development of four distinctive user personas. We discuss the design implications of these factors and personas and the potential that politeness holds for the design of conversational interfaces. The contributions of our work include:
\begin{enumerate}
    \item \emph{Design Space Exploration ---} We categorized voice interactions into speech acts and applied linguistic politeness strategies to the dialogue behaviors of a smart display application designed for older adults.
    \item \emph{Empirical Findings ---} The field study provided insights into the experience and preferences of older adults in using the smart display; their preferences regarding the features and interaction modalities offered by the smart display; the challenges they faced in the interaction; and their perceptions of the conversational agent.
    \item \modified{\emph{Design Recommendations ---} The empirical findings point toward relational, social, and cultural considerations as well as implementation constraints in using politeness for conversation design. The design implications of our work also highlight the importance of personalization in the use of proactive behavior, interaction pace, and the interplay between the visual and speech-based interaction.}
\end{enumerate}

\section{Related Work}

\subsection{Conversation Design for Voice Interfaces}
Prior research has proposed design guidelines and heuristics for voice user interface design with reference to factors in human's natural conversations. Grounded in \emph{Conversational Analysis}, \citet{moore2019conversational} suggested borrowing principles of how people design their dialogues with considering the three characteristics: \emph{Recipient Design}, \emph{Minimization}, and \emph{Repair}. Drawing inspiration from Grice's Cooperative Principle in pragmatic theory and Nielsen's usability heuristics for user interface design, \citet{10.1145/3411764.3445312} developed a set of usability heuristics to guide and evaluate the design of conversational agents. \citet{8383654} also suggested that the system ``[a]dapt agent style to who users are, how they speak, and how they are feeling'' in the heuristics. Despite the efforts towards building conversational agents in HCI research, the debate has been ongoing on whether voice interfaces should be conversational, natural or human-like \cite{10.1145/3290605.3300705, 10.1145/3411764.3445579, 10.1145/2858036.2858288, 10.7146/aahcc.v1i1.21316}. 

Besides design guidelines, prior work has studied user needs for voice interfaces. \citet{10.1145/3411764.3445445} explored user needs of competence, autonomy and relatedness applying Self-Determination theory and suggested three aspects that reflect user autonomy: control over the conversation, personalized experience and control over their data. Past research has identified two major purposes of voice assistant usage---social and transactional---and suggested different design strategies for each purpose \cite{10.1145/3411764.3445579, 10.1145/3411764.3445445, 10.1007/978-3-030-78635-9_58}. Prior work also revealed differences between people who have utilitarian orientation and those with relational orientation in their responses to a service robot's politeness during malfunction \cite{lee2011effect}.

\subsection{Linguistic Strategies in Conversation Design}

 Prior work has investigated the use of linguistic strategies in conversational interfaces and robots. \citet{6483608} used linguistic cues to increase the perceived expertise of an informational robot. \citet{lockshin2020we} suggested applying linguistic norms in human dialogue to autonomous agents, identified three key norms: directness, brevity, and politeness, and studied people's adherence to these norms under different contexts. Particularly, politeness as an linguistic strategy has been widely studied for autonomous agents, e.g., educational agent \cite{10.5555/1562524.1562568, 10.1016/j.ijhcs.2005.07.001}, recommendation systems for elderly care \cite{10.1007/978-3-319-31510-2_27}, repair for a malfunctioning service robot \cite{lee2011effect}, and help-seeking for a robot \cite{10.1145/2858036.2858217}.
 

An agent employing politeness can affect a user's perception of the agent, their sense of autonomy, and whether they think interacting with an agent is pleasurable. \citet{lee_influence_2017} found that robots in a healthcare setting were more likely to improve patient compliance when using polite behavior in their speech and gestures. \citet{kim_co-performing_2019} found that a user's initial mental model of the agent affected what information the user was willing to share and how they expected the information to be utilized. \citet{10.1007/978-3-030-78635-9_58} found that user preferences to disclose information changed based on how they perceived a chatbot that acted more task-oriented or more companion-like.

\subsection{Voice interfaces for older adults}
As the population of older adults grows worldwide, aging technologies are increasingly researched and built to improve senior people's quality of life. However, older adults often face challenges in using technology because of declines in cognitive (e.g., reduced memory), perceptual (e.g., hearing or vision), and motor (e.g., reduced dexterity) skills \cite{nurgalieva_systematic_2019}. Smart devices such as voice assistants and smart displays could address some of these challenges by providing older adults with multiple ways to interact with the device. Research has explored how older adults perceive and interact with voice assistants and has shown that they find voice assistants useful \cite{trajkova_alexa_2020, portet_design_2013}. However, older adults still face challenges using voice assistants because older adults don't fully understand how voice assistants work and/or how to use them \cite{koon_perceptions_2020, kim_exploring_2021, sayago_voice_2019, lopez_aging-focused_2019}. Likewise, voice assistants raise concerns about privacy and data collection due to their microphones continuously listening for wake words to activate the device \cite{trajkova_alexa_2020, kim_exploring_2021, lau_alexa_2018, manikonda_whats_2018}. Other issues, not specific to older adults, involve problems interacting with the voice interface. Users often encounter issues due to the device not activating, issues with the device performing the command, or the voice assistant misunderstanding the user's dialect or phrasing of a command. During an error, voice assistants may recover by giving a non-answer forcing the user to try the command or series of commands again \cite{zubatiy_empowering_2021}.

\section{Conversation Design}
\modified{To explore how linguistic politeness theory might inform the design of conversational interfaces that improve user experience while maintaining efficiency, we identified salient and repeated speech acts and matched politeness strategies to each act. We implemented these strategies into the conversational speech of a smart display application designed for older adults and conducted a formative study to assess the extent to which our designs reflected politeness. The paragraphs below provide further design on our conversation design approach.}
\subsection{Linguistic Strategies}

\begin{table*}[htb]
\caption{Conversational goals of and opportunities for the use of politeness strategies for speech acts.}
\small
\centering
\begin{tabular}{ p{6em} p{35em} p{18em} }

\toprule
\textbf{Speech Acts} & \textbf{Conversational Goal} &\textbf{Selected Politeness Strategy}\\
\midrule

\multirow{6}{*}{\textbf{Suggest}} & \multirow{6}{35em}{Protect user freedom of choice and minimize impositions; show care for the user; and be willing to help.}  & N1: be conventionally indirect\\
    & & N3: be pessimistic\\
    & & P6: avoid disagreement\\
    & & P7: presuppose common ground\\
    & & P12: include both S and H in the activity\\
    & & P14: assume or assert reciprocity \\
\midrule

\textbf{Comment} & Appreciate user effort and progress. & P15: gift\\

\midrule
\multirow{2}{*}{\textbf{Welcome}} & \multirow{2}{35em}{Show interest in the user and be inviting.} & P14: assume or assert reciprocity\\
& & P15: gift\\

\midrule
\multirow{2}{*}{\textbf{Farewell}} & \multirow{2}{35em}{Show interest in the user and be inviting.} & P6: avoid disagreement\\
& & P12: include both S and H in the activity\\

\midrule
\multirow{3}{*}{\textbf{Request}} & \multirow{3}{18em}{Minimize impositions of the request.} & N1: be conventionally indirect\\ 
& & N3: be pessimistic\\
& & P12: include both S and H in the activity\\

\midrule
\textbf{Repair} & Show regret and ask for forgiveness. & N6: apology\\

\midrule
\multirow{3}{*}{\textbf{Instruct}} & \multirow{3}{35em}{Be respectful and considerate; avoid impositions on user actions; and encourage user participation.} &  N1: be conventionally indirect\\
& & P7: presuppose common ground\\
& & P10: offer\\

\bottomrule
\vspace{6pt}
\label{tab:speech act goals}
\end{tabular}
\end{table*}

Language use is concerned with ``what is said'' and ``how what is said to be said'' \cite{grice1975logic}. Grice's Maxims describe ``how'' as the \emph{manner} dimension of cooperation, which emphasizes clarity, brevity, and order with the goal of effective information exchange. The Cooperative Principle, which includes manner, has been widely adopted by researchers and designers to design voice interfaces where brevity is encouraged to achieve the maximum efficiency \cite{10.1145/3411764.3445312}. On the other hand, \citet{brown_politeness_1987} proposed Politeness Theory as a deviation from Grice's Maxims. According to Politeness Theory, language usage has social meanings, and thus message construction is strategically used to express social relationships. Four strategies in conversation consider social needs: \textit{bald-on-record}, \textit{positive politeness}, \textit{negative politeness}, and \textit{off-record}. Bald-on-record follows Grice's Maxim of manner and is the most direct, clear, concise, and unambiguous way to communicate, whereas the politeness-based strategies conveyed messages in a more indirect way.

\modified{
Based on the two linguistic models, we designed a \textit{direct} and a \textit{polite} system. The direct system follows the manner from Grice's Maxims and the bald-on-record strategy from Politeness Theory. The polite system used linguistic strategies that minimized face-threatening acts, particularly the \textit{positive politeness strategy} that aims to promote the user's positive self-image and the \textit{negative politeness strategy} that focuses on the user's freedom of action. We categorized the system's utterances into seven speech acts: request, comment, suggest, repair, instruct, welcome, farewell.}

\modified{
To apply politeness strategies in a systematic and principled fashion, we first identified face needs for each speech act, as shown in Table \ref{tab:speech act goals}. For instance, the primary face need for the speech act \textit{request} is to minimize face threatening acts, such as reducing the impositions of the ask, while the primary need for \textit{comment} is to satisfy the user's desire to be liked, understood, and cared about. For each speech act, we identified the face needs of the act and matched them with politeness strategies that meet these needs. For example, to promote positive face needs when making comments, the system may say, ``I know it's not easy,'' using the positive politeness strategy ``presuppose/raise/assert common ground.'' The system may also comment with ``Well done'' to meet the hearer's need to be liked using the positive politeness strategy ``gift.'' Furthermore, interest can be promoted by the statement, ``Fantastic,'' which uses the strategy ``exaggerate.'' In making requests, the primary face need is to avoid negative face threatening acts where negative politeness strategies may be appropriate. For example, in making requests, the system can use ``may'' or ``can you,'' based on the strategy ``be conventionally indirect;'' can use ``could you'' and ''would you'' based on the strategy ''be pessimistic;'' and use ``just'' to adopt the strategy, ``minimize the imposition.'' 
}

\modified{
A mixture of positive and negative politeness strategies was also used to produce a hybrid strategy. According to Brown and Levinson \cite{brown_politeness_1987}, mixing strategies could have a softening effect. For instance, when making a suggestion, the system may say ``Let's take the chronic pain lesson. Maybe start now?'' Here, the positive strategy is the use of ``let's'' to include both the speaker and hearer in the activity in order to trigger the cooperative assumption. On the other hand, the use of ``maybe'' in this suggestive utterance is an example of negative politeness strategies that are conventionally indirect and that give the hearer an ``out'' for the suggestion. In this case, the face threatening effect of ``let's'' was softened by ``maybe,'' while ``let's'' remains to show inclusiveness and consideration for the hearer and indicates that the suggested action is for the user's benefit. Overall, positive politeness strategies were mostly applied to comment, welcome, and farewell, whereas negative politeness strategies were mainly used for repair and instruct. Table \ref{appendix: online_study_prompts} provides the full list of the strategies evaluated in the formative study. 
}



\subsection{System Details}
\subsubsection{System Features}
\modified{
We chose the Google Nest Hub Max as the smart display platform for our implementation and field evaluation. We adopted a health-intervention system, which has been shown to improve older adults' quality of life in prior research \cite{gustafson2021effect}}, to build a smart display application involving four key features: Discussion Groups, Chronic Pain Lesson, Health Survey, and Journal. Particularly, \emph{Discussion Groups} allows users to create public posts and comment on each other's posts. \emph{Chronic Pain Lesson} contains a series of educational videos on chronic pain and users can choose the resources beneficial for them to watch. \emph{Health Survey} asks users to reflect on their physical conditions and provide ratings for eight health-related statements. \emph{Journal} is a private blog for users to keep records of their thoughts and view their previous entries. 

\subsubsection{Voice Interface}
\begin{figure*}
    \centering
    \includegraphics[width=\textwidth]{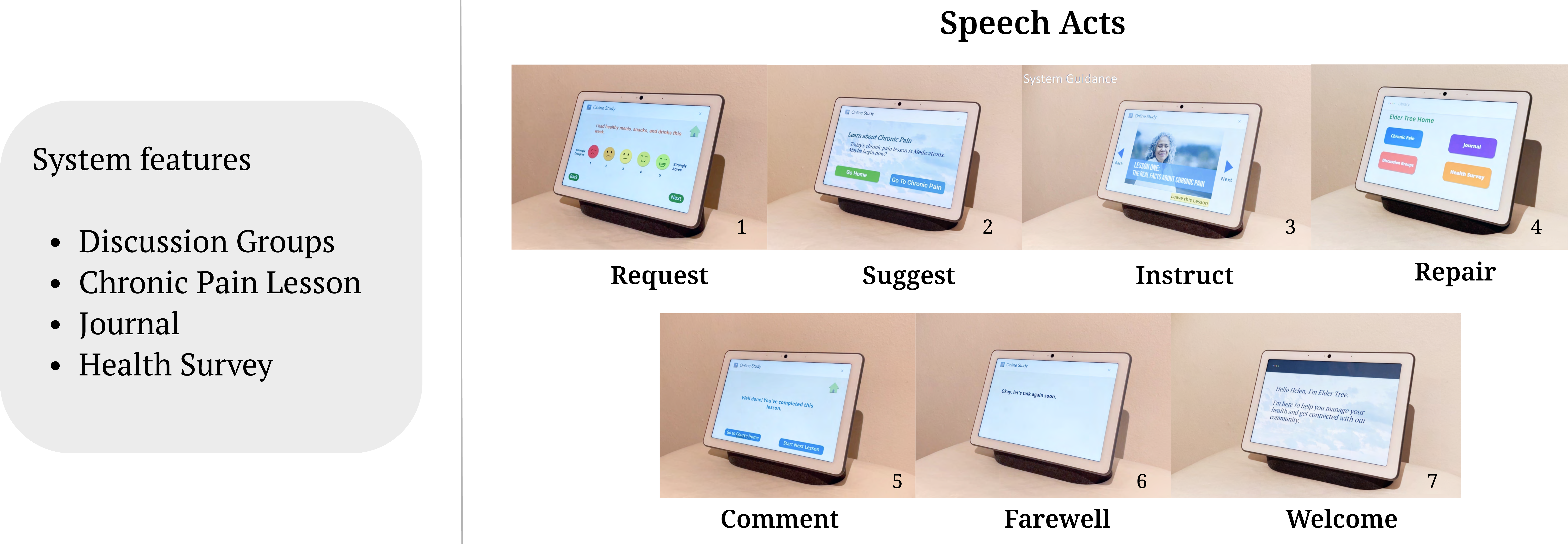}
    \caption{\modified{An overview of system features and examples of speech acts. (1) Making requests in \emph{Health Survey} (bald-on-record): ``Tell me your rating from one to five.'' (2) Suggesting educational content (positive politeness): ``Today's chronic pain lesson is Medications. Let's begin the lesson now?'' (3) Giving instructions for lesson navigation (bald-on-record): ``Change the lesson slides with the back and next buttons.'' (4) Repair voice recognition failure (negative politeness): ``Sorry, repeat what you said?'' (5) Commenting on lesson completion (positive politeness): ``Well done!'' (6) Farewell on exiting (positive politeness): ``Let's talk again soon.'' (7) Welcome when opening (positive politeness): ``Hello \textit{username},  I'm here to help you manage your health and get connected with our community.''}}
    \label{fig:features_speech_act}
\end{figure*}

\modified{We first scripted the voice based interaction flow for all four system features and categorized the utterances based on the seven speech acts (Figure \ref{fig:features_speech_act}). In particular, \emph{Discussion Groups} employs the speech acts suggest, request and instruct; \emph{Journal} includes suggest and comment; \emph{Chronic Pain Lesson} contains instruct and comment; and \emph{Health Survey} has request and comment. Moreover, the speech acts repair, farewell, and welcome were applied in the overall verbal interaction flow. }

\subsubsection{Touch Interface}
\modified{
We built a visual interface to support conventional, touch-screen-interaction-based control of the system. The visual screen served three purposes. First, the visual feedback informed the user about the current state of the system, such as when the system is processing the user's voice input and loading responses. Second, button text suggested terms that the system could understand. Third, clicking on the screen provided an alternative way to navigate the system, which sometimes proved more efficient, as it enabled the user to interrupt the system's speech. 
}

\subsubsection{System Implementation}
We used the framework Interactive Canvas of Google Actions to build both voice and touchscreen input and output. We implemented the voice interface with the Actions library and constructed a deterministic conversation model where all the voice prompts of the system were authored in advance. The conversation model was implemented with intents, types, and scenes, which predefined a set of linguistic concepts for the system to recognize and make responses.


\subsection{Formative User Study}
To inform design decisions, we conducted a formative study to evaluate the selected politeness strategies. We selected one speech example for each type and came up with a variety of utterances using each selected linguistic strategy. We created a total of 32 videos for the seven speech acts and each video used one strategy (7 videos for suggest, 4 for comment, 5 for request, 4 for repair, 3 for farewell, 5 for instruct, and 4 for welcome). One of the selected videos for each speech act was randomly assigned to the participant and each subject watched seven videos in the study. They were asked to evaluate the politeness of voice prompts with semantic differential scales. \modified{The scales were adopted from prior work on the use of politeness strategies in email composition \cite{lim2021designing}: directness of the speech (indirect to direct, ambiguous to straightforward), positive politeness (unfriendly to friendly, unsympathetic to caring), and negative politeness (demanding to undemanding, disrespectful to respectful).} Participants also evaluated their likeness of the prompt and willingness to follow with seven Likert scale questions. We included one attention check question asking the scenario of the previous video. We collected demographic information after they finished watching all seven videos. 

We used the crowdsourcing platform Amazon Mechanical Turk to conduct the online study. We approved 64 participants out of the recruited 70 workers ($ female=19, male=45 $), aged 19--71 ($M=36, SD=11.00$). The 6 responses were rejected due to the failed attention checks. Among all participants, 56 out of 64 had not used voice interfaces (87.5\%), and 14 of them had used one almost every day in the past month, suggesting that participants were familiar with voice interfaces.

Three outcomes of the online study have informed the system design. First, we identified the linguistic strategies with the highest politeness ratings for each speech type and used them more frequently in the final system. These strategies were verified to have higher ratings on politeness than the bald-on-record strategy. Second, we verified that the bald-on-record strategies for request, comment, welcome, farewell, repair achieved the highest score for directness in their category. Third, we identified the bald-on-record design with the highest directness scores. \modified{The evaluated linguistic strategies and results are shown in Table \ref{appendix: online_study_prompts}}.

\section{Evaluation}
\subsection{Hypotheses} 

We proposed the following two hypotheses on how the politeness of the system might affect older adults' overall experience using the system:

\begin{enumerate}
    \item [H1.] Participants will have a better overall experience with the polite system, using positive and negative politeness strategies, than the direct system using bald-on-record strategies.
    \item [H2.] Participants will be more tolerant of issues in the polite system than in the direct system.
\end{enumerate}

\subsection{Participants}
We recruited 16 participants over 60 years old in local communities. One user quit the study after the first two days of participation. Among the 15 participants ($ female = 8, male = 7$), who were aged 61--86 ($M=74.27, SD = 6.93$), none of them had prior experience with a smart display, \modified{while two users were experienced with smart speakers and had actively been using them prior to the study.} Participants self-evaluated their technological proficiency on a seven-point rating scale where seven is the most technologically savvy ($Min = 2, Max = 7, M = 4.93, SD = 1.44$). \modified{At the end of the study, each participant received \$20 USD in compensation.}

\subsection{Study Design}
\begin{figure}[!t]
    \includegraphics[width=0.5\textwidth]{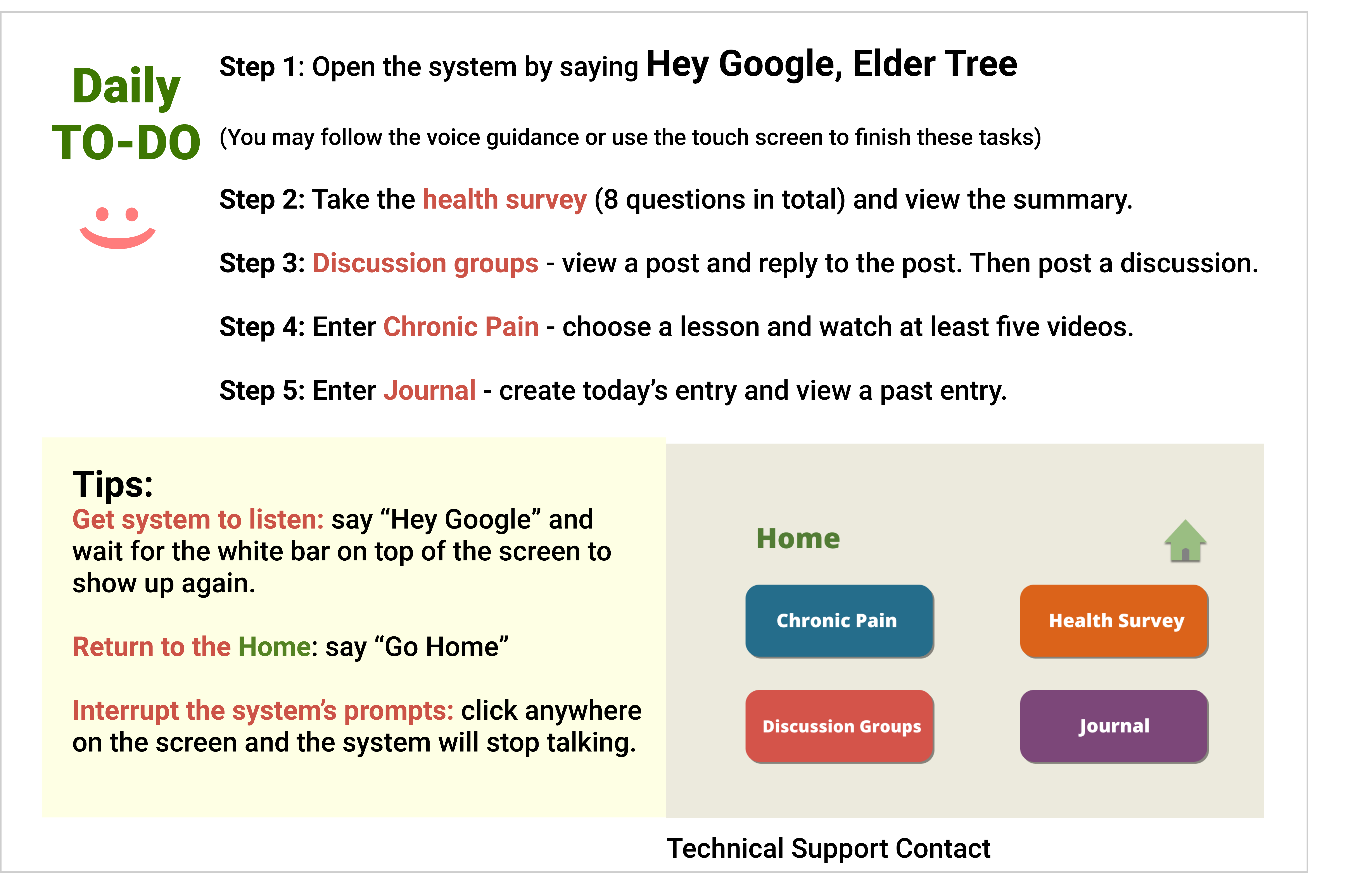}
    \caption{The cheat sheet with system guidance and tips of usage was provided for the participants during the intake appointment. Participants were asked to complete a daily to-do list following the steps on the cheat sheet.}
    \label{fig:cheatsheet}
\end{figure}

\paragraph{Study Setup}
\modified{We conducted a within-participants-design field study in an in-home environment in which participants used two versions of the conversational interface for a total of 10 days: (1) the \textit{polite} system that utilized positive politeness and negative politeness strategies, and (2) the \textit{direct} system that implemented the bald-on-record strategy. Participants experienced one condition for five days followed by the other condition for another five days. The order of the manipulation was randomized (8 of 15 participants used the polite version in the first session). Participants were asked to complete four daily tasks on the device following guidance on a provided cheat sheet (Figure \ref{fig:cheatsheet})}.

\begin{table}[!b]
\caption{Selected measures for the linguistic strategies with factor analysis results.}
\small
\centering
\begin{tabular}{ p{9em} p{18em} }
\textbf{Strategy} & \textbf{Items}\\
\toprule
\multirow{2}{9em}{Positive politeness ($\alpha = 0.70$)} & The system sounds friendly. \\ 
& The system sounds caring. \\ 
\midrule
\multirow{3}{9em}{Negative politeness ($\alpha = 0.71$)} & The system sounds demanding. \\ 
& I feel obligated to fulfill the system’s request. \\
& I am not in control of the system. \\
\midrule
\multirow{3}{9em}{Bald-on-record ($\alpha = 0.69$)} & The way the system communicates with me is efficient. \\
& The system sounds straightforward.\\
\bottomrule
\end{tabular}
\vspace{6pt}

\label{tab:factor_analysis_questionnaire}
\end{table}

\paragraph{Procedure}
\modified{Participants provided informed consent through email and scheduled in-person/remote intake sessions over the phone. To introduce the study and install the system, two experimenters visited the homes of 13 participants. The other two participants received the device by mail and completed the intake session remotely, as they were not local residents. The introduction provided participants with an overview of the voice and touchscreen controls of the smart display, the features of the application, and the daily tasks to be completed during the study. Following the intake appointment, participants completed the first five days of the study, followed by a questionnaire and semi-structured interview session. Next, the experimenters (remotely) switched the system version for the second session and notified participants of the change. At the end of the study, participants completed the same questionnaire and interview. Due to variation in participant availability and scheduling constraints, the study duration ranged from 9 to 13 days. Participants received \$20 USD at the end of the study. The study protocols were reviewed and approved by the appropriate institutional review board.}



\subsection{Measurement}
We collected three types of data to measure the effects of politeness strategies on people's experiences. 

\paragraph{System Usage} \modified{We calculated a number of measures of system usage based on system logs. These measures included the success rate of the interaction, user engagement, and user preference for input modality. The interaction success rate measure was computed as the percentage of completed tasks out of the four daily tasks assigned to participants. User engagement was measured by the interaction duration within the overall application and each feature on a daily basis. Finally, input modality preference was measured by calculating the number of user inputs through voice and touchscreen. }

\paragraph{Questionnaires} We measured the perceived system politeness, the experience of system failures, and the overall satisfaction and usability scores with three questionnaires. The first questionnaire asked users about their perception of the linguistic strategies. We conducted factor analysis to select the measures for positive politeness, negative politeness, and bald-on-record strategies. Cronbach's $\alpha$ values for these measures, shown in Table \ref{tab:factor_analysis_questionnaire}, indicated acceptable reliability. The second questionnaire asked about users' tolerance for system failures. The third questionnaire asked about their overall satisfaction and measured their overall experience using the System Usability Scale (SUS).

\paragraph{Thematic Analysis} We conducted semi-structured interviews to collect user feedback on system politeness, understand system expectations, and the overall experience. The interview questions had three topics. The first group asked users to reflect on their overall experience of using the application on the smart display. The second topic was related to system politeness, comparing the system's voice interaction with human conversations, and the perceived role of the device. The third topic asked about users' preferences over touchscreen and voice interactions for system inputs and outputs. We conducted thematic analysis on the interview data with two coders and summarized the common themes. 

\section{Results}
\subsection{Quantitative Results}
\begin{table*}[htb]
\caption{A summary of quantitative analyses of questionnaire responses. Normality (N) of the data was assessed using Shapiro-Wilk test. We used paired t-test if the difference is normally distributed and Wilcoxon test otherwise.}
\centering
\small
\begin{tabular}{rccccccc}
\hline
\textbf{Measures}                            & \textbf{Normality} & \textbf{Condition} & \textbf{mean} & \textbf{std} & \textbf{test type}                    & \textbf{statistic}                   & \textbf{p-value}                      \\ \hline
\multirow{2}{*}{Overall Likeness}          & \multirow{2}{*}{N} & Polite            & 5.23       & 1.35      & \multirow{2}{*}{wilcoxon test}   & \multirow{2}{*}{32.0}                & \multirow{2}{*}{0.54}  \\ \cline{3-5}
                                           &                    & Direct            & 5.07       & 1.46      &                                  &                                      &                                      \\ \hline
\multirow{2}{*}{Negative Politeness Score} & \multirow{2}{*}{N} & Polite            & 2.93        & 1.35      & \multirow{2}{*}{wilcoxon test}   & \multirow{2}{*}{45.5}                & \multirow{2}{*}{0.30}  \\ \cline{3-5}
                                           &                    & Direct            & 3.07       & 1.54      &                                  &                                      &                                      \\ \hline
\multirow{2}{*}{Directness Score}          & \multirow{2}{*}{Y} & Polite            & 5.17       & 1.40      & \multirow{2}{*}{paired t-test} & \multirow{2}{*}{-1.02} & \multirow{2}{*}{0.84}  \\ \cline{3-5}
                                           &                    & Direct            & 5.63       & 1.53      &                                  &                                      &                                      \\ \hline
\multirow{2}{*}{Positive Politeness Score} & \multirow{2}{*}{N} & Polite            & 6.17       & 0.84     & \multirow{2}{*}{wilcoxon test}   & \multirow{2}{*}{22.00}                 & \multirow{2}{*}{0.08} \\ \cline{3-5}
                                           &                    & Direct            & 5.97       & 0.99      &                                  &                                      &                                      \\ \hline
\multirow{2}{*}{System Error Score}        & \multirow{2}{*}{Y} & Polite            & 5.27       & 1.36       & \multirow{2}{*}{paired t-test} & \multirow{2}{*}{1.84}  & \multirow{2}{*}{$0.04^{*}$} \\ \cline{3-5}
                                           &                    & Direct            & 4.31       & 1.89      &                                  &                                      &                                      \\ \hline
\multirow{2}{*}{Satisfaction Score}        & \multirow{2}{*}{Y} & Polite            & 4.17       & 1.55      & \multirow{2}{*}{paired t-test} & \multirow{2}{*}{0.08}  & \multirow{2}{*}{0.47}  \\ \cline{3-5}
                                           &                    & Direct            & 4.14       & 1.81      &                                  &                                      &                                      \\ \hline
\multirow{2}{*}{SUS Score}                 & \multirow{2}{*}{Y} & Polite            & 69.17       & 13.81      & \multirow{2}{*}{paired t-test} & \multirow{2}{*}{0.99}  & \multirow{2}{*}{0.17} \\ \cline{3-5}
                                           &                    & Direct            & 64.83       & 20.34      &                                  &                                      &                                      \\ \hline
\multirow{2}{*}{Polite}                    & \multirow{2}{*}{N} & Polite            & 6.87       & 0.35     & \multirow{2}{*}{wilcoxon test}   & \multirow{2}{*}{6.00}                 & \multirow{2}{*}{$0.04^{*}$}  \\ \cline{3-5}
                                           &                    & Direct            & 6.67       & 0.62     &                                  &                                      &                                      \\ \hline
\multirow{2}{*}{Direct}                    & \multirow{2}{*}{Y} & Polite            & 6.20           & 1.15      & \multirow{2}{*}{paired t-test} & \multirow{2}{*}{-1.70} & \multirow{2}{*}{0.94} \\ \cline{3-5}
                                           &                    & Direct            & 6.60           & 0.63     &                                  &                                      &                                      \\ \hline
\end{tabular}
\label{tab:quant_result_questionnaire}
\end{table*}

\begin{table}[!b]
\caption{A summary of our quantitative result from system logging. In the repeated measures for system usage under two conditions, we used a linear mixed model to analyze the system logging data with conditions as fixed effects and subjects as random effects.}
\centering
\small
\begin{tabular}{rccccc}
\hline
\textbf{Measures} & \textbf{Cond.} & \textbf{LS Mean} & \textbf{St Err} & \textbf{F} & \textbf{p-value}  \\ \hline
Task Completion      & Polite            & 0.74       & 0.05        & \multirow{2}{*}{0.09} & \multirow{2}{*}{0.77}  
\\ \cline{2-4}
Rate & Direct            & 0.74       & 0.05      &  \\ \hline
Percentage of  & Polite            & 0.65        & 0.03        & \multirow{2}{*}{1.99} & \multirow{2}{*}{0.18}  
\\ \cline{2-4}
Voice Input & Direct            & 0.69       & 0.03      & \\ \hline
Interaction          & Polite            & 24.37       & 3.52      & \multirow{2}{*}{2.02} & \multirow{2}{*}{0.18}  
\\ \cline{2-4}
Duration (m/day) & Direct            & 19.43       & 3.52    &\\ \hline
Discussion & Polite            & 5.89       & 1.38     & \multirow{2}{*}{1.02} & \multirow{2}{*}{0.33} 
\\ \cline{2-4}
Duration (m/day) & Direct            & 5.10       & 1.38     &\\ \hline
Survey        & Polite            & 1.81       & 0.19     & \multirow{2}{*}{0.015}  & \multirow{2}{*}{0.91} 
\\ \cline{2-4}
Duration (m/day) & Direct            & 1.78       & 0.19      &\\ \hline
Lesson     & Polite            & 10.55       & 1.75      & \multirow{2}{*}{2.99}  & \multirow{2}{*}{0.11}  
\\ \cline{2-4}
Duration (m/day) & Direct            & 6.67       & 1.74      & \\ \hline
Journal                  & Polite            & 1.42       & 0.30   & \multirow{2}{*}{0.39}  & \multirow{2}{*}{0.55} 
\\ \cline{2-4}
Duration (m/day)   & Direct            & 1.25       & 0.30      & \\ \hline
\end{tabular}
\label{tab:quant_result_system_usage}
\end{table}


\modified{Informed by our hypotheses, all tests used one-tailed comparisons. The analysis of our questionnaire data and manipulation checks used paired t-tests or Wilcoxon tests, depending on normality. We analyzed our system usage data using linear mixed model and fit the model using the Restricted Maximum Likelihood Method (REML). We provide test details for marginal and significant effects in the text below and details for other effects in Tables \ref{tab:quant_result_questionnaire} and \ref{tab:quant_result_system_usage}.}

\modified{H1 hypothesized that people would have a more positive experience with the polite system than the direct system. The analysis of questionnaire responses showed no significant effects of the system's politeness on usability, likability of the voice interaction, or overall user satisfaction as shown in Table \ref{tab:quant_result_questionnaire}. Therefore, H1 was not supported. H2 hypothesized that people would be more tolerant of system errors in the polite condition than in the direct condition. However, the results showed an effect in the opposite direction (Table \ref{tab:quant_result_questionnaire}): participants were more tolerant of system errors in the direct condition ($M=4.31$, $SD=1.89$) than in the polite condition ($M=5.27$, $SD=1.36$), $t(14)=1.84$, $p=0.04$. Therefore, H2 was not supported. We further discussed the finding opposite to H2 in the Limitations Section.}

\modified{Besides hypotheses, we measured the effects of politeness strategies on system usage. Our analysis revealed that there was no significant effect on the task completion rate, system input modality, or interaction duration, as reported in Table \ref{tab:quant_result_system_usage}. Furthermore, we performed manipulation checks by evaluating people's perceptions of the system's politeness in both conditions. When asked to rate the statement ``The system sounds polite,'' participants provided a higher score for the polite system ($M=6.87$, $SD=0.35$) than the direct one ($M=6.67$, $SD=0.62$), $w(14)=0$, $p=0.04$. Although the result was significant, the actual differences are small and may not be meaningful. The results revealed no differences in participants' perceptions of the system's directness, negative politeness, or positive politeness, suggesting that participants were not able to differentiate among different politeness strategies.}

\subsection{Qualitative Results}
\begin{figure*}
    \centering
  \includegraphics[width=6 in]{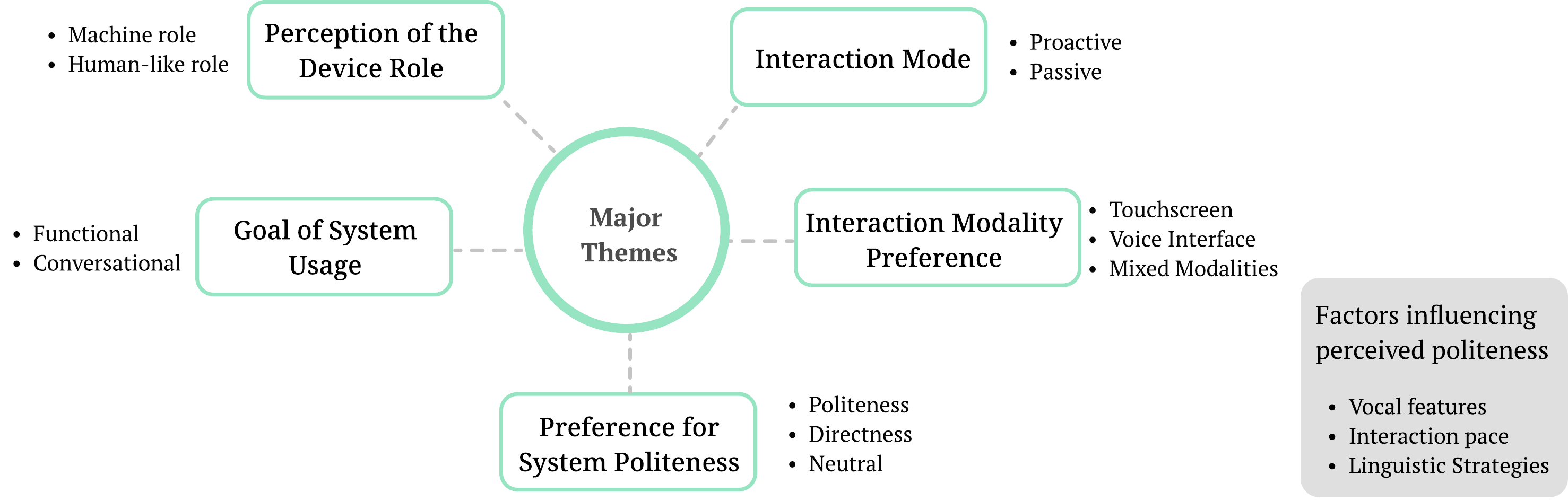}
  \caption{A diagram showing the major themes of our qualitative findings.}
  \Description{}
  \label{fig:themes}
\end{figure*}

Our qualitative findings merged into five themes in terms of user experience with the polite system and the direct systems as shown in Figure \ref{fig:themes}. 

\subsubsection{Factors that shape the politeness of the system }
\paragraph{Non-linguistic Features} Two non-linguistic factors that people attributed to shape the system's politeness were vocal features, such as tone, pitch and vocal inflection, and the pace of the interaction they had with the system. \modified{In both conditions}, participants described the vocal features as ``soft,'' ``low key,'' ``earnest,'' ``not so cheery'' and ``soothing.'' For the pace of the interaction, participants associated politeness with the slowness of the system's pace, thinking that the system ``was not pushing'' (\textit{D5}) and that ``it gives you ample time to make the decision before you answer, which is good. especially for some older folks, that it might take them a little longer to understand it'' (\textit{P3}).

\paragraph{Linguistic Features}

In both politeness conditions, participants pointed out phrases they thought were conventionally polite behaviors (\textit{P8, 11, 15; D7, 15}), such as ``it greets you,'' ``uses your name,'' ``thank you,'' and said ``nice to see you again.'' For negative politeness, participants described the system as being non-abrasive, non-confrontational, and not contentious, e.g., ``it prompts well for your options and prompts in a non-demanding way. It doesn't say press any key to continue or anything like that'' \modified{(\textit{P13})}. For positive politeness, users described the system as being friendly, kind, respectful, caring, considerate, understanding, and non-judgemental, e.g., ``it asks your question. You have a chance to answer. Doesn't say what is right or wrong or whatever'' \modified{(\textit{D5})}. For bald-on-record effects, people described the system as being clear, direct, and straightforward. They commented that ``the manner in which the questions were asked seemed very straightforward,'' and ``she speaks distinctly'' \modified{(\textit{D4; P3})}. Overall, the comments related to linguistic politeness are observed in both conditions. 

\subsubsection{Preference for system politeness}

\paragraph{Prefer Politeness}
Most people responded to the system's overall politeness in a positive way and described it as being pleasant and friendly. One user mentioned that ``I liked that idea that wasn't so robotic. It greets you'' \modified{(\textit{D11})}. However, the effects of linguistic politeness strategies were subtle and implicit for most users. They commented positively on the polite version while they couldn't tell the reason, saying that ``I think the second version was a little bit better. I'm not really sure why'' \modified{(\textit{P9})}. Only one user pointed out the linguistic strategy differences between the two versions and complained about the commanding manner of the direct version: ``I thought [it] said just pick the number then I'm like...that was my only beef with that one'' \modified{(\textit{D12})}.

\paragraph{Prefer Directness}
One participant disliked the system's politeness \modified{(\textit{P14})}, saying that ``I don't love being talked to by a computer as friendly as it sounds.'' She also viewed the linguistic polite strategy as ``added politeness'' and ``odd,'' saying that ``They weren't necessary, and they are just a little bit confusing.''

\paragraph{Neutral towards Linguistic Strategies}
Some participants had a neutral view regarding the system's politeness and thought that ``it was fine'' and they ``had no problem with it'' \modified{(\textit{P9, 14; D9})}. One user mentioned that the effects of politeness were related to their proficiency in using the system and ``after about the second or third time, it's just a voice'' \modified{(\textit{P2})}. 


\subsubsection{Goal of System Usage}

\paragraph{Functional Goals}
\modified{Some participants expressed goals that were more ``functional'' in nature toward completing study tasks}. These participants used the device to learn educational content or ask about the weather and news \modified{(\textit{P3, 6})}. Participants also experimented with the system in an attempt to ``mess it up'' or to find bugs in the system \modified{(\textit{P4, 8, 13; D13})}, e.g., ``I'm helping testing to help work out some of the problems. Not trying to get the benefit out of the content.'' \modified{(\textit{P8})} Finally, some participants said that their only goal was to complete the tasks on that checklist provided by the experimenter, e.g., ``I'm just doing some tasks... 'cause I was doing it for you'' \modified{(\textit{P15})}.

\paragraph{Conversation Goals} 
Some people tend to hold more ``conversational'' or ``relational'' goals when interacting with the system. They thought the system could help to reduce isolation for older adults, \modified{e.g., ``for those who don't have a lot of social interaction, I think this would boost their spirits tremendously'' (P12). These participants also commented on the enjoyment of the conversational experience, for example, stating ``it turns out to be a lot more pleasant than you expected'' (\textit{P10})}.


\subsubsection{Perception of Device Role}

\paragraph{Machine Role}
 \modified{In both conditions}, 11 out of 15 participants viewed the system as a machine, describing it as a device, AI, computer, or robot. They mentioned that the verbal interaction was systematic, mechanical, and unnatural. Similarly, they described the system as having a limited vocabulary, an unnatural interaction flow, and a lack of intellectual abilities. Regarding the limited vocabulary, participants mentioned that the communication content was strictly constrained as the system only understands limited commands, ``it could only say a certain number of things,'' \modified{(\textit{D14})} The unnatural interaction flow is reflected by the delayed system responses, not being able to interrupt the system's prompts, and missing meaningful social cues that ``you can see when you're making sense'' \modified{(\textit{P7})}. Furthermore, participants mentioned that the system lacked intellectual abilities, such as interesting ideas, humor, and expressing and reading emotions. 
 Users that viewed the system as a machine held mixed views towards the conversational nature of the voice interaction. Some users stated that there was no conversation at all, e.g., ``You have yes or no answers, so it really doesn't amount to a conversation'' \modified{(\textit{P14})}. Some users thought it was an unnatural conversation such as ``a formal conversation where you're limited to the words that you can use'' \modified{(\textit{D2})}.

\paragraph{Human-like Role}
 \modified{In both conditions}, the other four participants attributed diverse human roles to the system, calling it a friend, teacher, or professional. The perception that the system is a companion is associated with the understanding and caring characteristics of the speech and is described as ``a friend out in the universe...has my interests at heart'' \modified{(\textit{P12})} or ``interested in my views on things'' \modified{(\textit{D10})}. The professional and knowledgeable impression was described as ``some other person that knows what they're doing, professional'' \modified{(\textit{P5})}, which is mostly related to the system's educational content. Overall, users have attributed human roles to the system described it as being friendly and pleasant.

\begin{figure*}[!t]
    \centering
    \includegraphics[width=5.5 in]{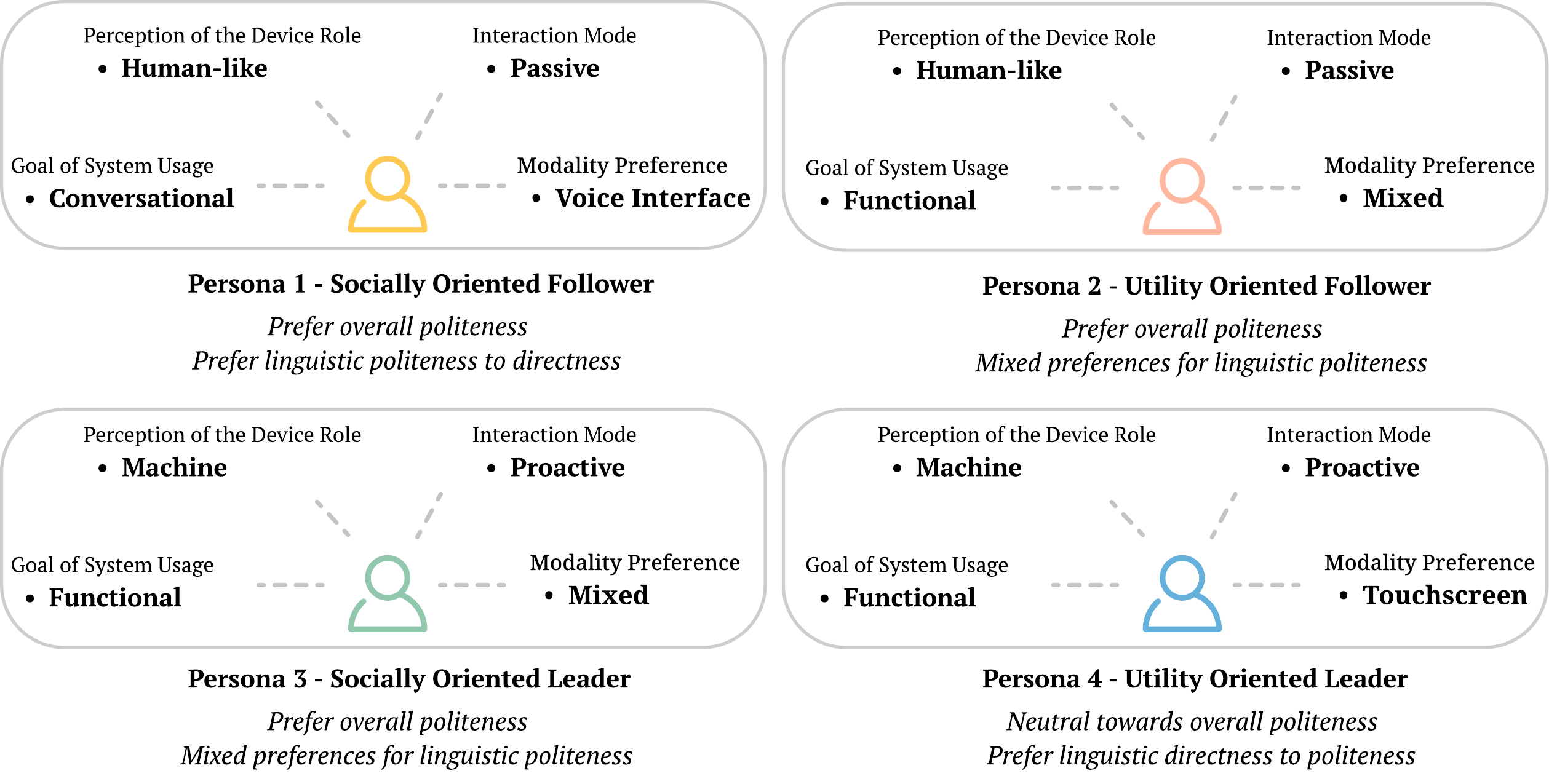}
    \caption{We identified four user personas shaped by their goals of interaction, preferences towards politeness, perceived device role and interaction modality preferences.}
    \label{fig:personas}
\end{figure*}

\subsubsection{Interaction Mode}
\paragraph{Proactive Interaction Mode} 
The proactive users preferred to take the lead in the interaction and seek information actively. For instance, two users mentioned that ``I tend to sort of look around for things rather than being guided to them'' \modified{(\textit{P13})} and ``I would rather do my own research, read about it than have it spoken to me'' \modified{(\textit{D14})}. Sometimes they would bypass the voice prompts by clicking the screen to speed up their interaction. Also, they tended to focus on the commands that the system could understand and thought that ``The terms you can use [are] obviously very limited'' \modified{(\textit{D13})}.

\paragraph{Passive Interaction Mode}
The more passive users preferred to be guided by the system and follow its requests. They tended to describe the system as being easy-to-follow and easy-to-understand, saying that ``I just gotta listen'' \modified{(\textit{D5})} and ``it directs you'' \modified{(\textit{D11})}. They preferred to hear the system's voice, saying that ``it's hearing another voice, you know giving you direction like reading the bedtime story to the kid before she goes to sleep'' \modified{(\textit{D12}) and ``the voice is friendlier than just reading it'' (\textit{P10})}.

\subsubsection{Interaction Modality} 
\paragraph{Prefer the Voice Interface}
We observed three main reasons for users' preference towards the voice interface: \modified{health conditions, distance of interaction and ease of use}. Some participants with vision or mobility issues can interact with the device remotely, e.g., one participant (\textit{P10}) explained that for people ``who don't see that well even with glasses'' the system has ``to have a voice.'' Furthermore, voice interaction with the device also allows for multi-tasking, e.g., ``I could be fixing a cup of tea or coffee...or taking my medications or putting them together'' \modified{(\textit{P11})}. Also, some people preferred the voice interface as it is ``easy to follow'' and ``sets everything at ease'' \modified{(\textit{P10, 12})}.

\paragraph{Prefer the Touchscreen}
The major reason that users preferred the touchscreen is to speed up the interaction and ``bypass the interactions just by tapping the screen'' \modified{(\textit{P14})}. Participants also commonly used the touchscreen to avoid or repair voice recognition failures. Some participants mentioned that they used the touchscreen because of their previous technology experience. Users may also prefer the touchscreen because of health conditions. One participant mentioned that he couldn't hear the device well because of a hearing issue \modified{(\textit{P4})}. 

\paragraph{Prefer the Mixed Modalities}
Some participants responded that they had gotten used to using the touchscreen in the first five days, but found the verbal commands to be easier to use so chose to use voice more \modified{(\textit{P5, 8})}. Some people preferred to have the option to switch the modalities based on the tasks at hand or their physical position relative to the device.

\subsection{User Personas}
We summarized four major personas representing the experience that older adults had with the smart display, as affected by their goal of system usage, perception of device role, interaction modes, and preference over system politeness (Figure \ref{fig:personas}).  

\paragraph{Persona 1---Socially Oriented Follower} The first user persona perceives the device as a friend and holds a conversational goal in the interaction \modified{(\textit{12})}, describing the interaction as ``communicating with someone on the outside'' (\textit{P12}). This persona tends to have a passive interaction mode and follow the system's guidance in exploring different features. This persona prefers politeness more than directness and has a positive experience with the system's overall politeness.

\paragraph{Persona 2---Utility Oriented Follower} The second persona holds a functional goal of using the system, perceives the device's role as a human and tends to follow the system's guidance \modified{(\textit{3, 5})}. Participants in this persona described the system as being ``knowledgeable'' and ``professional.''  \modified{(\textit{P3, 5})}. They ``follow what the operator said'' \modified{(\textit{D3})} and have positive feedback towards the system's overall politeness. This persona has mixed preferences towards the input modality, while prefers to hear the system's prompts rather than reading the screen, e.g., \modified{``I enjoyed, you know, listening to her'' (\textit{D3})}, while ``I just seem to automatically want to touch it'' \modified{(\textit{P3})}.

\paragraph{Persona 3---Socially Oriented Leader} Persona 3 contains users that have functional goals and perceive the device in a non-human role such as machine, AI, and robot \modified{(\textit{2, 4, 6-9, 11, 13})}. This persona is positive towards overall politeness but doesn't expect the system to be human-like, saying that ``I don't think the goal should be to give the voice personality beyond being polite and also being efficient. (\textit{P1})'' Users in persona 3 have a neutral or positive view towards the system's linguistic politeness.

\paragraph{Persona 4---Utility Oriented Leader} Similar to the third user persona, persona 4 holds functional goals and views the system as a machine \modified{(\textit{14})}. Compared with the other user personas, persona 4 uses the touchscreen the most often---whenever they can---and takes the lead in the interaction bypassing the voice guidance. This persona holds neutral or even negative views towards the overall politeness and prefers the linguistic directness rather than the linguistic politeness, describing the linguistic politeness as being ``odd'' and ``unnecessary'' (\textit{P14}).



\section{Discussion}

\modified{Our findings point to three areas where this work can inform the future design of smart displays for older adults: design for the politeness of conversational systems; design for both voice and touch modalities of smart displays; and personalization based on the four identified user personas. We discuss these areas below.}

\subsection{Design Implications for Politeness in Conversation Design}

\paragraph{Factors influencing perceived politeness}

Our findings revealed additional factors that shape user perceptions of the system's politeness. These factors highlight a rich design space for polite conversation design, including the manner that the message is delivered, the vocal features of the system's voice, and the pace of the interaction. Although our study focused on linguistic cues grounded in Politeness Theory \cite{brown_politeness_1987}, we observed several vocal features, including tone, pitch, reflection, and key of the voice, to contribute to perceived politeness. Politeness is also shaped by the pace of the interaction, as some users associated the slowness of the voice interaction with the system's politeness. \modified{As suggested by Eelen \cite{eelen2014critique}, politeness in interpersonal relationship ``can include non-verbal, non-linguistic behaviour,'' such as nodding and turn-taking in a conversation, and even ``the absence of behaviour,'' e.g., returning greeting with silence, which can be seen as being impolite. In the same vein, the politeness of a conversational agent can manifest itself through both linguistic cues and non-linguistic features, and the politeness design of smart systems should consider both opportunities.}

\paragraph{Design with Politeness Strategies}
\modified{Brown and Levinson \cite{brown_politeness_1987} viewed interaction as ``the expression of social relationships built out of strategic language use'' in which politeness is tied with relationship-building between speaker and addressee. The design of polite speech should also consider the relationship between the user and the device. Our study showed that people attributed both machine and human-like characteristics to the system, which may suggest different status relationships between the user and the device, also coined as ``relative power'' \cite{brown_politeness_1987}. For instance, people who view the system as a tool may assume a higher power status, and the use of positive politeness strategy to establish common ground may backfire, as they thought the ``added politeness...weren't necessary, and they are just a little bit confusing'' (\textit{P14}). On the other hand, people who treat the system as ``a friend from the universe'' (\textit{P12}) may assume a similar power status, and the use of the politeness strategies could meet the user's expectations that they are in a conversation with a caring friend. In addition, some participants mentioned that they also thought about the people who built the system when talking to the device, suggesting that the manufacturer, designer, or the researcher involved in the development of the system may be stakeholders in the relationship between the user and the device. The ''relative power'' between these stakeholders and the user could also affect people's perceptions of the device's politeness. 
}
 
\modified{
In addition to the ``relative power'' between the user and the device, culture is another factor believed to impact the use of politeness strategies. Prior research suggests that people's polite behaviors are closely tied to social norms and cultural expectations \cite{ide1989formal, blum1984requests, gu1990politeness, eelen2014critique}. While cultural factors are not within the scope of the current study, we highlight the importance of situating the design of the system's polite speech in its cultural context. Certain speech acts in one culture may be interpreted differently in another, and polite behaviors should be culturally appropriate \cite{blum1984requests}. In addition to perceived politeness, culture can also affect user expectations of the device's politeness. Therefore, the design of polite speech should consider both the social meanings of the speech acts and the social acceptance of polite speech.
}

\modified{
The implementation of politeness strategies should also consider the system's technical capabilities. For instance, implementing \textit{off-record} strategies would require a high level of understanding of the user and the environment, e.g., the use of metaphors and association clues. Often the hearer needs to take the perspective of the speaker to disambiguate meaning conveyed through off-record strategies. Messages that use metaphors can cause confusion, as users need to take the perspective of the device, e.g., the perceived knowledge and experience of the device, to be able to correctly interpret the metaphor. While the current study focused on speech acts using \textit{on-record} strategies, the successful implementation of off-record strategies could potentially increase the perceived cognitive capabilities of conversational systems and the user's conversational experience with them.
}

\subsection{Design Implication for Smart Display Interaction}

\modified{
Our findings showed that participants have adopted a mixed modality of system control and output since the frequencies of using the voice interface and touchscreen were similar in our system usage data. Therefore, interaction design on smart displays should consider the distinctive features of each modality as well as the interplay between them. We offer three key design implications, which are discussed below.}

\modified{First, user feedback on the overall interaction suggests that the system's verbal design should consider both the system's vocal features and interaction flow guided by the system's prompts. The voice interaction could be categorized into two types: (1) dialogue for guiding the interaction and (2) audio for conveying information. In designing the verbal exchanges for system guidance, we categorized the system's utterances into speech acts and applied linguistic strategies by considering the face needs associated with each speech act. The analysis on system usage duration revealed that participants spent a considerable amount of time talking to the system each day, compared to simple queries such as asking about the weather and playing music. Therefore, it is important to support extended conversational exchanges through the voice interaction design.
}

\modified{
Second, the results suggest that participants commonly used the touchscreen to bypass voice outputs to improve interaction efficiency as well as handle errors when voice recognition failed. Therefore, the interaction on the touchscreen should provide shortcuts for faster system navigation and support the repair of interaction failures. Moreover, the visual display on the touchscreen should provide language cues on what verbal commands the system might understand and what the user could say to the system as suggested by Yankelovich \cite{10.1145/242485.242500}. 
}

\modified{
Third, the interaction modality switch that we observed among users illustrates that interaction design should consider the flexibility of choosing modalities to meet user needs under different conditions. For example, users might change modalities based on their physical distance from the device. We observed that some users placed the device on the dinner table and sat in front of it during the interaction, while others placed the device on a coffee table and interacted with it from a chair at some distance. Interaction distance can also be affected by which task the user is completing. For instance, users need to be close to the device to watch lesson videos, while they can answer survey questions remotely without looking at the screen. Users may also switch interaction modality while multitasking, e.g., one user mentioned that she had to talk to the device while making coffee or cleaning the house. Finally, we observed that some participants experiences physical limitations, such as loss of vision, hearing, and mobility, which made it difficult for them to hear the voice or see the device screen clearly. Hence, the system should allow for modality-switching at any time during the interaction to support user needs and the context of interaction.
}

\subsection{Design Implications for Personalization}
The four personas identified in our analysis are potential targets for personalization in terms of system politeness and the interplay between the voice and touchscreen modalities.


\paragraph{Design Opportunities for Socially Oriented Follower}

The focus on the conversational experience of \textit{Socially Oriented Follower} suggests that the system should be proactive to progress the interaction and guide users in the navigation. The finding also suggested that the tone, pitch, reflection, and gender of the voice should be personalized based on user preferences. The system should avoid being commanding or pushy and should leave sufficient time for users to think and respond to system prompts. With its potential to build a relationship with the user through conversation, this system could be particularly beneficial for older adults facing isolation and could more broadly offer mental health support.

\paragraph{Design Opportunities for Utility Oriented Follower}
To support \textit{Utility Oriented Follower}'s functional goal of information seeking, we suggest that the design should balance the conversational experience and the efficiency of system usage. The touchscreen should support all user inputs as an alternative for voice inputs, while the button text could be more conversational and less command-like, using words people would say to answer the system verbally, e.g., ``OK, let's begin'' instead of ''Begin.'' 

\paragraph{Design Opportunities for Socially Oriented Leader}
Given the positive feedback to system politeness from \textit{Socially Oriented Leader}, the conversation design should leverage politeness, utilizing both vocal features and linguistic politeness strategies, to improve the overall user experience. Furthermore, to improve interaction efficiency, the system could avoid confirmations or reading out options for a selection if it can confidently recognize that the user is viewing the screen. As with all adaptive systems, these features must be designed carefully to minimize errors due to a lack of proper recognition of user context.

\paragraph{Design Opportunities for Utility Oriented Leader}
\textit{Utility Oriented Leader} tended to treat the smart display no differently than a tablet or other device for most of its features. We therefore suggest that the interaction design makes use of linguistic directness to maximize efficiency. Because this user type tends to lead the interaction, the system should have a low level of proactivity and wait for user input to progress the interaction. 


\subsection{Limitations}
This work has a number of limitations that limit the applicability of our findings to other settings and that inform future work. First, our analysis of quantitative data showed that, contrary to our hypothesis, participants were marginally more tolerant toward errors by the more \textit{direct} version of the system over the more \textit{polite} version. A potential explanation of this finding is a mismatch between user perceptions of the system's capabilities and the errors it made. Specifically, polite speech might have been seen as being more human-like, suggesting human-like capabilities, and the errors that users observed in simple tasks, such as voice recognition violated this expectation. Future exploration of the specific effects of mismatch between language use and capabilities can provide more conclusive explanations of our observation, as recent research found the perceived competence of the system has an important effect on user experience \cite{koelsch2021impact}. 
\modified{Second, our evaluation of the politeness strategies in the formative study mainly focused on the functionality of speech acts without considering individual differences and cultural factors. Future studies may consider additional factors, such as cultural and social-contextual factors, that affect how speech acts are interpreted.} Furthermore, although our development adopted an iterative approach with continuous user input, we have observed a number of usability issues in our system that caused voice recognition failures, long loading times for touchscreen pages, and occasional termination of the smart display application. Although such issues are prevalent even in commercial smart display applications, they might have affected the overall user experience and our interpretation of the effects of the linguistic strategies we explored.

\section{Conclusion}
Conversational interfaces are becoming pervasive technologies integrated into smart speakers and display devices. These interfaces increasingly rely on human-like dialogue to offer a natural experience, which holds tremendous potential for older users. However, dialogue-based interfaces can be cognitively demanding and make error recovery difficult for this population of users. In this paper, we explored the use of Politeness Theory \cite{brown_politeness_1987} in designing conversations for a smart display to address some of these challenges toward improving user experience. We categorized the voice interaction offered by a smart display application designed for older adults into seven major speech acts---request, suggest, instruct, comment, welcome, farewell, and repair---and applied politeness strategies to each speech act. A field deployment study asked 15 older adults to use a more \textit{direct} or a more \textit{polite} version of a smart display for five days at their homes. From our data, we identified five factors related to preferences toward politeness, conversational interaction, and experiential orientation and characterized four user personas shaped by these factors. These factors and personas inform how politeness cues and strategies might be used by the future design of voice interfaces and point to opportunities for personalization.
\begin{acks}

This study is funded by the Agency for Healthcare Research and Quality, United States Department of Health and Human Services (grant number 1R18HS026853-01A1), received ethical approval from the University of Wisconsin Health Sciences and Minimal Risk Research Institutional Review Board (reference number 2020-0868), and is registered at ClinicalTrials.gov (NCT04798196). We would like to thank our collaborators at the Center for Health Enhancement Systems Studies (CHESS) at the University of Wisconsin--Madison for their assistance in the testing and deployment of our system and the administration of our field study. 

\end{acks}

\balance
\bibliographystyle{ACM-Reference-Format}
\bibliography{literature-review}

\appendix
\setcounter{table}{0}
\renewcommand{\thetable}{A\arabic{table}}

\section{Formative Study}
\subsection{Measures for Politeness Strategies - Semantic Differential Scale}
\begin{enumerate}
    \item Indirect---Direct.
    \item Ambiguous---Straightforward
    \item Unfriendly---Friendly
    \item Unsympathetic---Caring
    \item Demanding---Undemanding
    \item Disrespectful---Respectful
\end{enumerate}

\subsection{Measures for Overall Likeness of the Speech - Seven Likert Scale Questions}
\begin{enumerate}
    \item I like the way the system talks to me.
    \item I am willing to follow up with the system's prompts.
\end{enumerate}
We evaluated the selected politeness strategies for seven speech acts and created 32 videos playing the utterances on the smart display with visual feedback. The strategy and speech examples are summarized in Table \ref{appendix: online_study_prompts}.
\subsection{Formative Study - Evaluated Politeness Strategies}

Table \ref{appendix: online_study_prompts} provides the prompts used in our formative study to evaluate politeness strategies.

\section{Questionnaire to Evaluate the Polite and the Direct System}
The post-study survey has 35 Likert scale questions. For each question, we asked users to provide a rating from 1 to 7, where 7 is strongly agree and 1 is strongly disagree.

\noindent\textbf{Overall Likeness}
    \begin{enumerate}   
        \item Overall, I like the way the system talks to me.
        \item The conversation with the system is a positive experience.
    \end{enumerate}
    \textbf{Politeness}
    \begin{enumerate}
        \item The system sounds polite.
        \item The system sounds friendly.
        \item The system sounds caring.
        \item The system sounds distant to me.
        \item The system sounds disrespectful.
        \item The system sounds demanding.
        \item I feel obligated to fulfill the system's request.
        \item I feel hesitant to say no to the system.
        \item I am in control of the system.
        \item The way the system communicates with me is efficient.
        \item The system sounds direct.
        \item The system sounds assertive.
        \item The system sounds straightforward.
    \end{enumerate}
    \textbf{Tolerance of system issues}
    \begin{enumerate}
        \item It annoys me when it takes a long time for the system to load. 
        \item It annoys me when the system’s voice recognition fails. 
        \item It annoys me when unexpected system errors occur.
    \end{enumerate}
    \textbf{Satisfaction}
    \begin{enumerate}
        \item I am satisfied with it.	 
        \item I would recommend it to a friend.	 
        \item It is fun to use.
        \item It works the way I want it to work.	 
        \item It is wonderful.
        \item I feel I need to have it.	 
        \item It is pleasant to use.
    \end{enumerate}
    \textbf{SUS (System Usability Scales)}
    \begin{enumerate}
        \item I think that I would like to use this system frequently. 
        \item I found the system unnecessarily complex. 
        \item I thought the system was easy to use.  
        \item I think that I would need the support of a technical person to be able to use this system.  
        \item I found the various functions in this system were well integrated.  
        \item I thought there was too much inconsistency in this system. 
        \item I would imagine that most people would learn to use this system very quickly. 
        \item I found the system very cumbersome to use. 
        \item I felt very confident using the system. 
        \item I needed to learn a lot of things before I could get going with this system.  
    \end{enumerate}
    \label{tab: questionnaie_appendix}


\begin{table*}[htb]
\caption{\modified{Result summary of the formative study that informs the politeness strategy selection for each speech act. The scores are the mean for bald-on-record, positive politeness and negative politeness from left to right. We used a curated set of strategies that received relatively high politeness scores or that enriched language use. The selected bald-on-record, positive politeness, or negative politeness strategies are highlighted with $^*$, $\dagger$, and $\ddagger$, respectively.}}
\vspace{-4pt}
\centering
\resizebox{\textwidth}{!}{
\normalsize
\begin{tabular}{p{15em} p{35em} p{15em}}
\toprule
\textbf{Linguistic Strategies} & \textbf{Prompts} & \textbf{ Scores}                                                                                                                            \\ \toprule
\textbf{Suggest}               &                                                                                                                                                                          \\ \midrule
bald-on-record$^*$                           & Today's chronic pain lesson is Medications. Begin now?                   &             5.95/5.70/5.85                                                                                                        \\
bald-on-record                            & Today's chronic pain lesson is Medications. Don't miss it!  Begin now?     & 5.81/5.41/5.59                                                                                                                     \\

N1$\ddagger$                           & Today's chronic pain lesson is Medications. Maybe begin now?      &6.44/6.11/6.33                                                                                                                                                                     \\
N3$\ddagger$                            & Today's chronic pain lesson is Medications. Would you like to begin now? &6.19/6.13/6.00  
\\
N6$\ddagger$                             & Today's chronic pain lesson is Medications. Begin the lesson then?     &5.46/4.88/5.19                                                                                                                  \\
P12$\dagger$                            & Today's chronic pain lesson is Medications. Let's begin the lesson now? & 5.81/5.44/5.81                                                                                                                                    \\
P14$\dagger$                             & Here is a chronic pain lesson for you. Today's topic is Medications. Begin the lesson now?   & 6.60/5.80/5.50                                                                                                                                                                                                 \\
\midrule
\textbf{Comment}               &                                                                                                                                                                          \\ \midrule
bald-on-record$^*$                          & Lesson completed.       &6.50/5.24/5.83                                                                                                                                                     \\
P7                             & I know it takes time, but you completed this lesson! &6.13/5.26/5.48                                                                                                                                     \\
P15$\dagger$                             & Congratulations! You've completed this lesson. & 6.38/6.27/6.58                                                                                                                            \\
P15$\dagger$                            & Well Done! You've completed this lesson.   & 6.05/5.95/6.25                                                                                                                                       \\ \midrule
\textbf{Request}               &                                                                                                                                                                          \\ \midrule
bald-on-record$^*$                            & Tell me your rating from one to five. & 6.07/5.21/5.29                                                                                                                                     \\
P15                            & What is your rating from one to five? I\'m listening.  & 6.13/5.34/5.22                                                                                                                       \\
N1$\ddagger$                             & You may give me a rating from one to five.   & 6.42/5.92/5.79                                                                                                                          \\
N3$\ddagger$                             & Could you choose a rating from one to five? &    6.38/5.58/5.88                                                                                                                  \\
N4$\ddagger$                             & Just choose a rating from one to five.     & 6.55/5.30/5.05                                                                                                                              \\ \midrule
\textbf{Repair}                &                                                                                                                                                                          \\ \midrule
bald-on-record$^*$                            & Repeat what you said?             &6.11/5.39/5.39           \\
N6$\ddagger$                            & Sorry, could you repeat what you said?    & 6.00/5.36/5.54                                                                                                                          \\
N7                           & It seems that my voice recognition didn't work well. Repeat what you said? &  5.85/5.56/5.00                                                                                                 \\
N7                             & It was hard to catch what you said. Repeat what you said? & 6.03/5.63/5.58                                                                                     
                                                                                                \\ \midrule
\textbf{Farewell}              &                                                                                                                                                                          \\ \midrule
bald-on-record$^*$                            & Okay, bye. &	6.43/5.09/5.43                                 \\
P6$\dagger$                             & Okay, I'll talk with you next time then. &	6.00/5.95/5.73                                                                                                               \\
P12$\dagger$                            & Okay, let's talk again soon. &	5.83/5.70/6.07                                                                                                                                                                                                                                                                                                                                            \\ \midrule
\textbf{Instruct}           &                                                                                                                                                                          \\ \midrule
bald-on-record$^*$                            & Change the lesson slides with the back and next buttons. &	6.29/5.29/5.15 
\\
N1$\ddagger$                             & You may click the next or back buttons to change lesson slides. & 5.92/5.46/5.92                                                                                                           \\
N7$\ddagger$                            & The next or back buttons could be clicked to change lesson slides   & 6.40/5.57/6.07
\\
P10$\dagger$                            & I'm able to change lesson slides if you click the back and next buttons. & 6.55/5.35/5.90
\\
P12                            & Let's change lesson slides with the back and next buttons.      & 6.35/5.35/5.60                                                                                                                                                                                                                                                                                                  \\ \midrule
\textbf{Welcome}               &                                                                                                                                                                          \\ \midrule
bald-on-record                            & Welcome! & 5.61/5.24/5.76                                                                            \\
bald-on-record$^*$                           & Hello \textit{username},  welcome to \textit{application-name}! & 6.55/5.41/5.91                                                    \\
bald-on-record	 &	Hello \textit{username}, see what's new on \textit{application-name}!                 &	5.91/5.73/5.95                                                                                         \\
P14$\dagger$                             & Hello \textit{username}, I'm \textit{application-name}. I'm here to help you manage your health and get connected with our community.        &    6.20/5.83/5.80                                                                           
\\ \bottomrule
\end{tabular}}

\vspace{1ex}
\footnotesize
{\raggedright $\dagger$\textbf{Positive Politeness Strategy} P2: exaggerate; P6: avoid disagreement; P7: presuppose/raise/assert common ground; P10: offer, promise; P12: include both S and H in the activity; P13: give or ask for reasons; P14: assume or assert reciprocity; P15: gift \\
$\ddagger$ \textbf{Negative Politeness Strategy} N1: be conventionally indirect; N3: be pessimistic; N4: minimize the imposition; N5: give deference; N6: apology; N7: impersonalize
\par}
\label{appendix: online_study_prompts}
\end{table*}


\end{document}